\documentclass[twocolumn,showpacs,preprintnumbers,amsmath,amssymb]{revtex4}

\usepackage{graphicx}
\usepackage{dcolumn}
\usepackage{bm}
\usepackage{graphics}
\usepackage{amsmath}
\usepackage{amssymb}
\usepackage{amscd}
\usepackage{afterpage}
\usepackage{float,times}
\usepackage{subfigure}
\usepackage{rotating}
\usepackage{multirow}
\usepackage{fancyheadings}
\usepackage{epsfig}
\usepackage{theorem}
\usepackage{moreverb}
\usepackage{euscript}
\usepackage{psfrag}

\begin{document}

\title{Suppressing Proton Decay By Separating Quarks And Leptons}

\author{A. Coulthurst}\email{a.coulthurst@physics.unimelb.edu.au}

\author{K. L. McDonald}\email{k.mcdonald@physics.unimelb.edu.au}

\author{B. H. J. McKellar}%
 \email{b.mckellar@physics.unimelb.edu.au}
\affiliation{%
School of Physics, Research Centre for High Energy Physics, The
University of Melbourne, Victoria, 3010, Australia\\
}%

\date{\today}

\begin{abstract}
Arkani-Hamed and Schmaltz (AS) have shown that proton stability need not
originate from symmetries in a high energy theory.
Instead the proton decay rate is suppressed if quarks and
leptons are spatially separated
in a compact extra dimension. This separation may be achieved by
coupling five dimensional fermions to a bulk scalar field with a
non-trivial vacuum profile and requires relationships between the
associated quark and lepton Yukawa couplings. We hypothesise that these
relationships are the manifestation of an underlying symmetry.  We further
show that the AS proposal may suggest
that proton stability \emph{is} the result of an underlying
symmetry, though not necessarily the traditional baryon number symmetry. 
\end{abstract}

\maketitle

In recent years it has been proposed that the fundamental scale of
nature may be
much less than the Planck
scale~\cite{Arkani-Hamed:1998rs,Antoniadis:1998ig,Antoniadis:1990ew}. By introducing
large extra dimensions one is able to reframe the hierarchy problem
and remove the need to explain the disparity between the Planck scale
and the electroweak scale. In the standard model (SM) it is known that
proton decay proceeds at the non-renormalizable level via the dimension six operator $Q^3L/\Lambda^2$, where
$Q$ ($L$) generically denotes a quark (lepton) field operator and
$\Lambda$ is the SM cut-off. The
stringent lower bound of $1.6\times10^{33}$ years on the decay mode
$p\rightarrow e^+\pi$ leads to the bound
$\Lambda\gtrsim10^{16}$~GeV~\cite{Nath:2006ut}. In models with large
extra dimensions the fundamental gravitational scale may be reduced to
TeV energies, removing the order
$10^{16}$~GeV cut-off required to suppress the proton decay rate.

Arkani-Hamed and
Schmaltz (AS)~\cite{Arkani-Hamed:1999dc} have suggested that proton longevity
need not imply a conserved symmetry in the more fundamental theory~\cite{note_no_flava}. They have shown that proton decay can be suppressed in models with a low fundamental
scale if quarks and leptons are localised
at different four dimensional slices of a five dimensional
spacetime. If the fifth dimension forms an
$S^1/Z_2$ orbifold, maximal suppression of the proton decay rate
results when quarks and leptons are localised at different fixed
points. It is known that the zero mode of a five
dimensional fermion, which may be identified with a SM fermion, can be
localised at an $S^1/Z_2$ orbifold fixed
point by
coupling the fermion to a bulk scalar
field with a non-trivial vacuum
profile~\cite{Georgi:2000wb}. The sign of the associated Yukawa
coupling determines the fixed point at which the fermion zero mode is
localised~\cite{Georgi:2000wb,Kaplan:2001ga}. Thus proton decay may be
suppressed by arbitrarily choosing different sign Yukawa couplings for
quarks and
leptons with a bulk scalar (see e.g.~\cite{Lillie:2003sq}).

It is interesting to speculate that an underlying theory may possess
symmetries which fix the relative bulk scalar Yukawa coupling signs
between quarks and leptons. In this work we ask if the
separation of quarks and leptons required to
achieve the AS proposal may itself be
the manifestation of an underlying symmetry. Thus proton stability
would result from the symmetries of an underlying theory, though
not necessarily the traditional baryon number symmetry. Indeed, it was suggested
in~\cite{Arkani-Hamed:1999za} that
it may be possible to understand the separation of quarks and leptons
in a five dimensional $SO(10)$ model, through fermion couplings to a symmetry
breaking vacuum expectation value (VEV) in the $B-L$
direction. However this does not ensure that the theory will
separate quarks and leptons. This may be seen as follows. Consider a
five dimensional spacetime with the fifth dimension forming an
$S^1/Z_2$ orbifold. Take $\Psi$ as a bulk field in the
$\mathbf{16}$ of $SO(10)$, containing a family of SM fermions, and $H$ as a
bulk scalar in the adjoint representation of $SO(10)$. The
Yukawa Lagrangian for $H$ is
\begin{eqnarray}
\mathcal{L}_H=\sum_ig_i\bar{\Psi}_i\Psi_iH,\label{orb_pdecay_so10_lagrangian}
\end{eqnarray}
where $i=1,2,3$ labels the different generations. If $H$ develops a
kink profiled
VEV in the $B-L$ direction, chiral zero mode fermions will be localised
at one of the orbifold fixed points, with the point of localisation
determined by the sign of the Yukawa coupling between the given
fermion and the $B-L$ direction scalar $H_{B-L}$.
Observe that the 
quarks of a given
generation will couple to
$H_{B-L}$ with a different Yukawa coupling sign than the leptons of the same
generation. However quarks and leptons of different generations may
still couple to $H_{B-L}$ with the same sign. The mixing observed in the
quark sector requires all quarks to be separated from the light leptons
in order to suppress the proton decay rate. This will not occur unless
one arbitrarily
chooses $g_i>0$ or $g_i<0$ for all $i$. Thus breaking $SO(10)$ in the
$B-L$ direction by a bulk scalar does not guarantee
suppression of the proton decay rate in a higher dimensional theory.

In this brief note we assume that $(a)$ the hierarchy problem tells us
that the SM
cut-off must be low (order $\sim10$~TeV) and that $(b)$ the longevity of the
proton results from the separation of quarks and leptons in an extra
dimension. Desiring simplicity we further assume that $(c)$ the separation
of quarks and leptons results from the simplest Yukawa Lagrangian
which naturally localises quarks and
leptons at opposite boundaries of an $S^1/Z_2$ orbifold. We
identify a minimal set of symmetries required to preserve the Yukawa
coupling relationships this Lagrangian contains. In order to construct
a complete
theory possessing this minimal set of symmetries one is required to
extend the SM gauge group. We identify candidate extensions. Let us
emphasise that our main point is that \emph{proton stability may be the
manifestation of an underlying symmetry in the AS proposal}. We
illustrate this with a concrete example. Although alternative symmetries
which achieve quark-lepton separation may exist, our observation
holds independent of the specific construct. We note that neutrino mixing has recently been investigated using symmetrical configurations of bulk fermions~\cite{Dienes:2006bu}.

We assume a five dimensional product spacetime $M^4\times
S^1/Z_2$, where $M^4$ denotes a four dimensional Minkowski
spacetime. The action of the $Z_2$ transformation is defined by
$y\rightarrow -y$, where $y$ labels the extra dimension. We include
the following five dimensional fermions:
\begin{eqnarray}
U,D,N,E,U^c,D^c,E^c,N^c,\label{orb_pdecay_fermions}
\end{eqnarray}
where the zero modes of the fields $U,D,N,E$ form the
usual SM $SU_L(2)$ doublets and the zero modes of $U^c$, $D^c$ and
$E^c$ will be identified with
the SM $SU_L(2)$ singlet fields. We have included a SM gauge singlet
field $N^c$ for reasons which will
become evident. The zero mode of this field is
the charge conjugate of the usual right-chiral neutrino. We also
include a gauge singlet
bulk scalar $\Sigma$. The action of the orbifold discrete symmetry
$Z_2$  on the fields is
\begin{eqnarray}
\Sigma(x^\mu,y)&\rightarrow& \Sigma(x^\mu,-y)=-\Sigma(x^\mu,y),\nonumber\\
F(x^\mu,y)&\rightarrow& F(x^\mu,-y)=\gamma^5F(x^\mu,y),\label{orb_p_decay_orbifold_bc}
\end{eqnarray}
where $x^\mu$ labels $M^4$, $F$ generically
labels the fermions~(\ref{orb_pdecay_fermions}) and $\gamma^5$ is the
usual product of Dirac matrices. The scalar potential
is given by
\begin{eqnarray}
V(\Sigma)=\frac{\lambda}{4\Lambda}(\Sigma^2-u^2)^2,\label{orb_pdecay_scalar_potential}
\end{eqnarray}
where $\lambda$ is dimensionless and $\Lambda$ is the cut-off. The combination of the scalar's orbifold parity and the potential $V$
result in the VEV profile
~\cite{Kaplan:2001ga}
\begin{eqnarray}
\langle\Sigma\rangle(y)&\approx&u\tanh[\beta
y]\tanh[\beta(L/2-y)],
\end{eqnarray}
where $\beta^2=\lambda Lu^2/4$ and the orbifold fixed points are
located at $y=0$ and $y=L/2$. The fermion orbifold parities (\ref{orb_p_decay_orbifold_bc})
permit only the (four dimensional) left-chiral component of a given
fermion $F$ to possess
a zero mode (call it $f_L^{(0)}$). The sign of the Yukawa coupling constant between the
field $F$ and $\Sigma$ then determines the fixed point at which this
zero mode $f_L^{(0)}$ is localised~\cite{Georgi:2000wb}.

In general, the Yukawa Lagrangian for $\Sigma$ takes the form
\begin{eqnarray}
-\mathcal{L}_{\mathrm{Yuk}}=\sum_{F,i}
h_{F_i}\overline{F}_iF_i\Sigma,\label{orb_pdecay_general_yukawa_lagrangian}
\end{eqnarray}
where the sum is over all fermion fields
(\ref{orb_pdecay_fermions}) and all generations $i=1,2,3$. The Yukawa constants $h_F$ are in general
independent and the separation of quarks and leptons required to
ensure proton longevity demands that one
enforce the relationships
\begin{eqnarray}
\mathrm{sign}(h_q)=-\mathrm{sign}(h_l),\label{orb_pdecay_yukawa_relationships}
\end{eqnarray}
where the subscript $q$ ($l$) labels quarks (leptons). We shall not
employ the most general Yukawa Lagrangian
(\ref{orb_pdecay_general_yukawa_lagrangian}). Instead we assume the simplest Yukawa Lagrangian which naturally
separates quarks and leptons, namely
\begin{eqnarray}
-\mathcal{L}_{\mathrm{Min}}&=&h\left\{\sum_i(U_i^2 + D_i^2+U_{i}^{c2}+ D_{i}^{c2})\right.\nonumber\\
& &\left.-\sum_i(N_i^2+E_i^2+N_{i}^{c2}+E_{i}^{c2})\right\}\Sigma,\label{orb_pdecay_minimal_lagrangain}
\end{eqnarray}
where $h$ denotes a common Yukawa
coupling constant and we ignore quark colour for the moment. We employ an obvious notation with $F^2=\bar{F}F$. The sign of $h$ determines the fixed point at which, e.g.,
quarks are localised, with leptons automatically localised at the opposite boundary
of the compact extra dimension.

The simplicity of (\ref{orb_pdecay_minimal_lagrangain})
motivates us to identify five
dimensional extensions of the SM which naturally produce this Yukawa
Lagrangian. To this end we note that $\mathcal{L}_{\mathrm{Min}}$ possesses the
symmetry~\cite{cont_sym_footnote_orb_pdecay}
\begin{eqnarray}
\mathcal{G}=U(3)_f\otimes U(4)_g\otimes Z_2^{QL}.\label{orb_pdeacy_symmetry_L_min}
\end{eqnarray}
Here $U(3)_f$ $\left[U(4)_g\right]$ is the group of unitary rotations of three [four] objects and $Z_2^{QL}$ is
a discrete symmetry interchanging two objects. The sets of
fermions
\begin{eqnarray}
\left\{U_{i},D_{i}, U_{i}^{c},D_{i}^{c}\right\}\mkern10mu\mathrm{and}\mkern10mu
\left\{N_{i},E_{i},N_{i}^{c},E_{i}^c\right\}\label{orb_pdecay_reps_yuk_lagrangian}
\end{eqnarray}
each form $(3,4)$ representations of $U(3)_f\otimes U(4)_g$, with $U(3)_f$
mixing the three families and $U(4)_g$ mixing the four states
within a family. The symmetry $Z^{QL}_2$ is defined by the
interchange of the two sets
(\ref{orb_pdecay_reps_yuk_lagrangian}). Note that $\Sigma$ transforms
trivially under $U(3)_f\otimes U(4)_g$ and is necessarily odd under $Z_2^{QL}$. 

Clearly the group $\mathcal{G}$ is not a symmetry of the SM and
naively we may expect any
extension of the SM which reproduces the Lagrangian
$\mathcal{L}_{\mathrm{Min}}$ to exhibit this rather restrictive symmetry. There are, however, more transparent subgroups of
$\mathcal{G}$ which, when enforced upon an extended SM,
ensure the Yukawa coupling relationships in
$\mathcal{L}_{\mathrm{Min}}$. First observe that 
\begin{eqnarray}
U(4)_g\supset Z_2^{LR},
\end{eqnarray}
where the action of $Z_2^{LR}$ is defined by
\begin{eqnarray}
U_{i} &\leftrightarrow&U_{i}^{c},\nonumber\\
D_{i} &\leftrightarrow&D_{i}^c,\nonumber\\
N_{i} &\leftrightarrow&N_{i}^c,\nonumber\\
E_{i} &\leftrightarrow&E_{i}^{c}.
\end{eqnarray}
Also note that
\begin{eqnarray}
U(3)_f\supset Z_3^{f},
\end{eqnarray}
where $Z_3^f$ is a cyclic symmetry group acting on the three
generations of fermions. The symmetry group
\begin{eqnarray}
\mathcal{G}_{\mathrm{Min}}=Z_3^f\otimes Z_2^{LR}\otimes Z_2^{QL},
\end{eqnarray}
is in fact a minimal symmetry set required to preserve the Yukawa
coupling relations in $\mathcal{L}_{\mathrm{Min}}$.

Why is the group $\mathcal{G}_{\mathrm{Min}}$ more transparent than
$\mathcal{G}$? The groups $Z^{LR}_2$ and $Z_2^{QL}$ are the
familiar discrete symmetries found in the Left-Right (LR)
symmetric model and the Quark-Lepton (QL) symmetric model
respectively. These models are well known and have been studied in
both four
dimensional~\cite{Pati:1974yy,Mohapatra:1974hk,Mohapatra:1974gc,Senjanovic:1975rk,Foot:1990dw,Foot:1991fk,Levin:1993sq,Shaw:1994zs,Foot:1995xx}
and 
higher dimensional~\cite{Mimura:2002te,Mohapatra:2002rn,Perez:2002wb,Mohapatra:2002ug,McDonald:2006dy}
frameworks. The LR model arises when one postulates that the
fundamental theory describing nature
possesses a left-right interchange symmetry. One is required to
introduce the additional gauge symmetry $SU_R(2)$ to permit
$Z_2^{LR}$. The QL model
results from the assumption that nature displays a quark-lepton
interchange symmetry at a fundamental level. When quark colour is
introduced in (\ref{orb_pdecay_minimal_lagrangain}) one must also introduce leptonic colour $SU_l(3)$ to permit
the QL symmetry.
The so
called Quark-Lepton Left-Right (QLLR) symmetric
model, which contains the discrete symmetry group $Z_2^{LR}\otimes
Z_2^{QL}$, has also been studied~\cite{Foot:1989wt}. Each of these models
would allow one to achieve some degree of quark-lepton
separation in 5D, but would not naturally separate all quarks and leptons.

An existing model
of greater interest to us is that based on the
gauge group~\cite{Foot:1990um,Foot:1990un}
\begin{eqnarray}
\mathcal{H}\equiv[SU(3)]^2\otimes
[SU(2)]^2\otimes[U_X(1)]^3.
\end{eqnarray}  
Here the
$SU(3)$ factors are the colour groups of the QL symmetric model, namely the
usual colour group $SU_c(3)$ and the leptonic colour group
$SU_l(3)$ (required to construct a QL symmetric Lagrangian). The $SU(2)$
factors are the familiar chiral groups of the LR model, whilst the fermion
quantum numbers under $[U_X(1)]^3$ distinguish the different
generations. Of importance is the fact that the model admits
the discrete symmetry group
\begin{eqnarray}
Z_3^f\otimes Z_2^{LR}\otimes Z_2^{QL},
\end{eqnarray} 
namely $\mathcal{G}_{\mathrm{Min}}$. Thus models based on the gauge group
$\mathcal{H}$ automatically admit the Yukawa Lagrangian
$\mathcal{L}_{\mathrm{Min}}$ in 5D and thus naturally separate quarks
and leptons. Having made this identification a few comments are in
order.

First note that the group $\mathcal{H}$ requires the
introduction of leptonic colour
$SU_l(3)$ in order to admit the discrete
symmetry $Z_2^{QL}$. It is known however that leptonic colour
does not emerge from many of the popular grand unified theory gauge groups like
$SU(5)$, $SO(10)$ and $E_6$. Nonetheless there exist alternative approaches
to gauge unification. Quartification models are based on the gauge group
$[SU(3)]^4$~\cite{Joshi:1991yn,Babu:2003nw,Demaria:2005gk,Demaria:2006uu}
and
admit the discrete symmetry $Z_2^{LR}\otimes
  Z_2^{QL}$. A 5D quartification model would allow one to
 obtain quark-lepton separation with three independent bulk scalar
 Yukawa coupling
constants (one per generation). In this sense the longevity of the
proton may be as natural in 5D quartification models as it is in the
proposal of~\cite{Arkani-Hamed:1999za}.

The fact that $Z_3^f\subset U(3)_f$
suggests that a unified model
containing a horizontal gauge symmetry would naturally explain the longevity
of the proton via quark-lepton separation. Quintification models offer
an alternative unified framework and employ the gauge group
$[SU(3)]^5=[SU(3)]^4\otimes SU_H(3)$~\cite{Ma:2005qy}, where the
subscript $H$ labels a horizontal symmetry. Thus it would be
possible to realize $\mathcal{L}_{\mathrm{Min}}$ within a five
dimensional quintification model, demonstrating that our approach may be
compatible with the notion of
unification.

In conclusion, we have investigated symmetries
which allow one to naturally separate quarks and leptons in five dimensional
models. This separation is important in that it allows one to
understand the long lifetime of the proton in models with a low
fundamental scale. We have shown that higher dimensional extensions of
the SM with the gauge
group $[SU(3)]^2\otimes
[SU(2)]^2\otimes[U_X(1)]^3$ admit
the discrete symmetry $Z_3^f\otimes Z_2^{LR}\otimes
Z_2^{QL}$ and allow one
to achieve quark-lepton separation in both a natural and minimal
fashion.

It is intriguing that in this approach proton
longevity remains a manifestation of underlying
symmetries in the high energy
theory, though not necessarily baryon number as
traditional approaches would suggest. Irrespective of the specific
example we have constructed, the observation that proton stability
within the AS proposal may imply underlying symmetries remains of interest.
\section*{Acknowledgements}
This work was supported in part by the Australian Research
Council. The authors thank R. Volkas, A. Demaria and D. George for
comments on the manuscript. 


\begin{thebibliography}{999}
\bibitem{note_no_flava} AS also sought to address flavour with their
proposal. We focus on proton stability in this
work and do not attempt to address flavor.
\end{thebibliography}
\end{document}